\documentclass[aps,pre,twocolumn,groupedaddress,showpacs,floatfix,superscriptaddress,longbibliography]{revtex4-2}
\usepackage{amsmath,amssymb}
\usepackage{physics}
\usepackage{graphicx}
\usepackage[dvipsnames,usenames]{xcolor}
\usepackage[normalem]{ulem}
\usepackage{soul}  
\usepackage[hyperindex,breaklinks,colorlinks=true, citecolor = blue]{hyperref}
\emergencystretch=\maxdimen
\hypersetup{
 colorlinks=true,
 linkcolor=blue,
 anchorcolor = blue,
 citecolor = blue,
 filecolor = blue,
 urlcolor = blue
}

\newcommand{\vx}{\pmb{x}}

\newcommand{\su}{\sigma^u}
\newcommand{\sigmad}{\sigma^d}

\newcommand{\CA}{ cellular automata }

\newcommand{\sutdphys}{Science, Mathematics and Technology Cluster, Singapore
University of Technology and Design, 8 Somapah Road, 487372 Singapore}
\newcommand{\oist}{Okinawa Institute of Science and Technology Graduate University, Onna-son, Okinawa 904-0495, Japan}

\newcommand{\sutdepd}{EPD Pillar, Singapore University of Technology and Design, 8 Somapah Road, 487372 Singapore}

\newcommand{\hnu}{Key Laboratory of Low-Dimensional Quantum Structures and Quantum Control of Ministry of Education, Department of Physics and Synergetic Innovation Center for Quantum Effects and Applications, Hunan Normal University, Changsha 410081, China
}
\newcommand{\cqt}{Centre for Quantum Technologies, National University of Singapore 117543, Singapore} 
\newcommand{\majulab}{MajuLab, CNRS-UNS-NUS-NTU International Joint Research Unit, UMI 3654, Singapore}

\begin{document}
\title{Tensor-Networks-based Learning of  Probabilistic Cellular Automata Dynamics} 

\author{Heitor P. Casagrande}
\email{heitor-peres@oist.jp}
\affiliation{\sutdphys}
\affiliation{\oist}

\author{Bo Xing}
\affiliation{\sutdphys}

\author{William J. Munro}
\affiliation{\oist}

\author{Chu Guo}
\affiliation{\hnu}

\author{Dario Poletti}
\email{dario\_poletti@sutd.edu.sg}
\affiliation{\sutdphys}
\affiliation{\sutdepd}
\affiliation{\cqt}
\affiliation{\majulab}

\begin{abstract} 
Algorithms developed to solve many-body quantum problems, like tensor networks, can turn into powerful quantum-inspired tools to tackle problems in the classical domain. 
In this work, we focus on matrix product operators, a prominent numerical technique to study many-body quantum systems, especially in one dimension. It has been previously shown that such a tool can be used for classification, learning of deterministic sequence-to-sequence processes and of generic quantum processes. We further develop a matrix product operator algorithm to learn probabilistic sequence-to-sequence processes and apply this algorithm to probabilistic cellular automata. This new approach can accurately learn probabilistic cellular automata processes in different conditions, even when the process is a probabilistic mixture of different chaotic rules. In addition, we find that the ability to learn these dynamics is a function of the bit-wise difference between the rules and whether one is much more likely than the other. 
\end{abstract}
  
\maketitle

\setcounter{figure}{0}


\section{Introduction} 

Machine learning has been widely and successfully used in different scientific fields and beyond \cite{alphago, chatgpt, CarleoZdeborova2019, schuld2015, adcock2015, biamonte2017} due to the broad expressive power and generality of the different architectures available. 
Between these architectures, tensor networks have also been recently considered as a version of quantum-inspired machine learning models \cite{Stoudenmire, Ran2020, Cheng2021, Shi2022}. They were first used for classification tasks and as auto-regressive generative models \cite{ HanZhang2018,stoudenmire2018, Liu_2019, Cheng2019, Sun2020, Liu2023, Ho_2023}. Under the name of exponential machines, they have also been applied for natural language processing tasks like sentiment evaluation \cite{deng2018}. 
In these examples, the tensor network architecture is usually one or two-dimensional \cite{Novikov, pestun2017}. Other architectures, like tree-tensor networks, have also been considered for tasks such as high-energy physics \cite{FelseMontangero2021} and radiology \cite{CavinatoMontangero2021}. 

Tensor networks have two significant advantages. The first one is that being linear, they can be trained very effectively compared to non-linear models. The second is that they can be efficiently compressed, reducing the number of parameters associated with the network in a controlled manner, while providing a controlled degree of accuracy \cite{schollwock2005, mcculloch2007, vidal2004}. 
However, tensor network models suffer from the inability to describe efficiently systems with significant long-range correlations. For instance, in one-dimensional quantum systems, they cannot efficiently describe states with volume-law entanglement \cite{SchollwockMPS, fannes1992, perez2006, verstraete2008, orus2014}.

In \cite{GuoPoletti2018} it was shown that tensor networks can be used to learn the deterministic evolution of sequences. In particular, the authors analyzed the evolution of cellular automata  \cite{Wolfram1983}, which is an ideal test bed because the dynamics can be very rich, yet the rules that generate it are simple and local.    
The authors showed that the tensor networks were able to distill the simple local dynamical rules from time-evolved \CA sequences.
Similarly, tensor networks were applied to learn generic, non-Markovian, quantum processes \cite{GuoChuMPSlearningNonMarkovian}, with important applications in quantum computing. 
In this work, we aim to move beyond deterministic evolution and consider probabilistic ones. In other words, we focus on the more general class of sequence-to-probabilistic-sequence scenarios, relevant for realistic noisy or probabilistic systems such as those encountered in biology \cite{louis2018}, or quantum mechanics \cite{PeresCasagrandePoletti2023}, to name just a few.
In the spirit of understanding in detail the functioning of this learning model, we consider the prototypical problem of learning probabilistic \CA \cite{louis2018}. More specifically, we consider the cases in which the dynamics include two or three different probabilistic rules, drawn from a given distribution. We show that our model can accurately learn the probabilistic \CA dynamics, regardless of the complexity of the underlying rules (e.g. regular or chaotic). We also find that the predictions are more accurate when the rules are similar bit-wise, and when their probabilities of occurrence are comparable. 

The rest of the work is organized as follows: in Sec.~\ref{sec:pcas} we give an introduction to probabilistic \CA. In Sec.~\ref{sec:MPOs} we introduce the basics of tensor networks and show the roles matrix product states and matrix product operators play in representing probabilities and conditional probabilities. In Sec.~\ref{sec:training} we explain how we prepare the data and train our model. In Secs.~\ref{sec:prediction} and~\ref{sec:characterization} we describe how to perform predictions and characterize the performance of the trained model. We provide analytical insights on the tensor network's ability to represent probabilistic \CA in Sec.~\ref{sec:analytical}. Our results are presented in Sec.~\ref{sec:results}. Lastly, we draw our conclusions in Sec.~\ref{sec:conclusions}.

\section{Probabilistic Cellular Automata}\label{sec:pcas} 
Given an input sequence $\vx = \left(x_1, x_2, \dots x_L\right)$ of length $L$ and $x_l \in (0,1)$, one can consider simple local rules to obtain an output sequence $\vx^{\prime}$. This type of system is known as \CA . In one dimension, there are $2^8=256$ local rules for which the value of $x^{\prime}_l$ depends on the values of $x_{l-1}$, $x_{l}$, and $x_{l+1}$ \cite{Wolfram1983}. These simple rules can result in attractive or periodic dynamics. Some rules are proven to be chaotic or even Turing complete~\cite{TuringCompleteCA, 10.1007/11786986_13}. 

\begin{figure}[ht]
\includegraphics[width=\columnwidth]{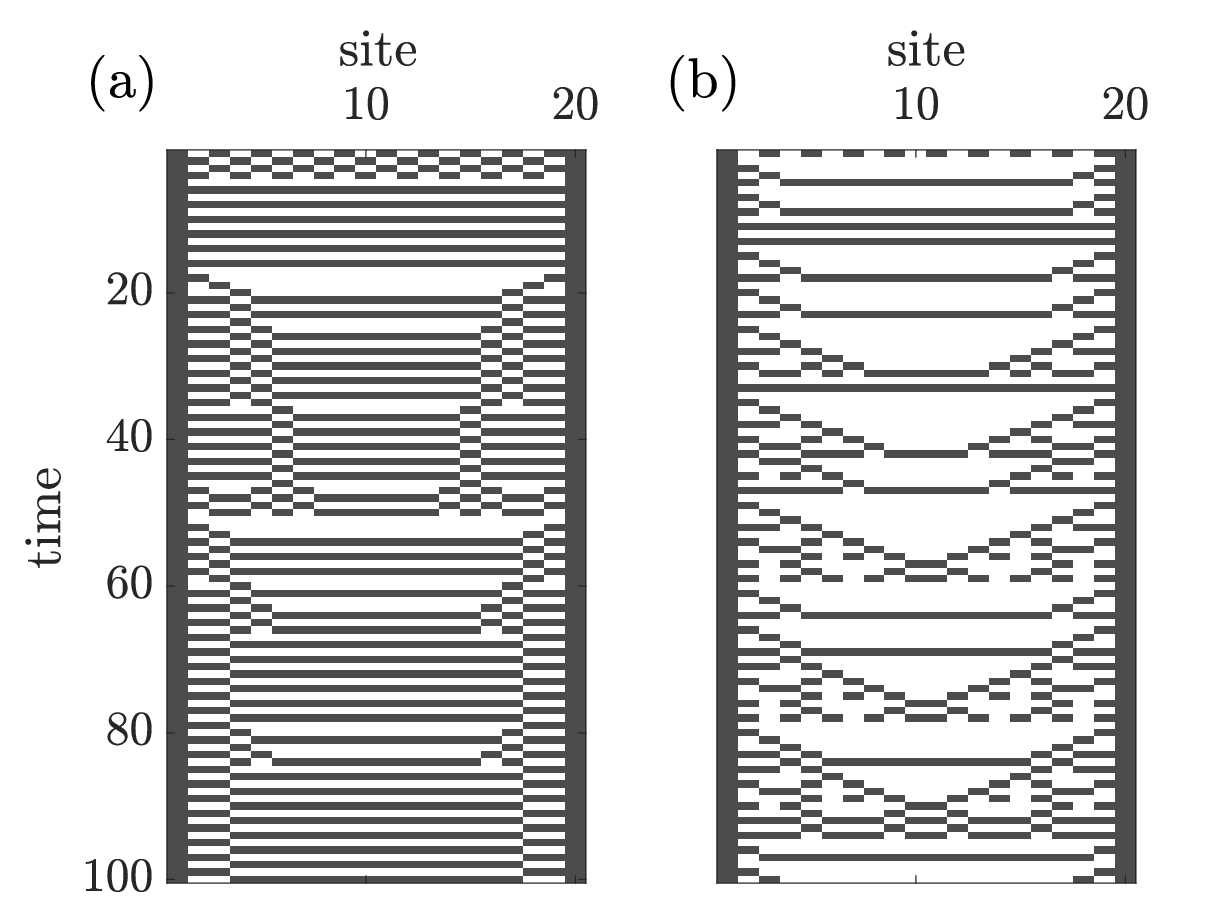}
\caption{Illustration of a single realization of a probabilistic \CA evolution between two different rules: rule 18 with probability $p_{18}$ and rule 51 with probability $p_{51}=1-p_{18}$. The left panel presents probability $p_{18} = 0.2$ while the right panel presents $p_{18} = 0.8$. At each time step, the sequence $\vx$ is updated to different $\vx^\prime$ following the two probabilistic \CA rules.}
\label{fig2:trajectories}
\end{figure}

In probabilistic \CA , the $\vx \to \vx^\prime$ dynamics depends on multiple local rules, each with a predefined probability of occurrence.
In a case where the dynamics depend on two different rules, one could apply rule $i$ with probability $p_{i}$ or rule $j$ with probability $p_{j}=1-p_{i}$. At each step,  a random number $r$ is drawn from a uniform, continuous distribution between $0$ and $1$. If $r\le p_{i}$, rule $i$ is applied. Otherwise, rule $j$ is applied. Cases with more possible rules can be dealt with using the same approach. 
Two examples of probabilistic \CA are visualized in Fig.~\ref{fig2:trajectories}. In panels (a) and (b), we apply rules 18 and 51 with different probabilities and observe vastly different dynamics.

\section{Matrix product operators and learning}\label{sec:MPOs}

Our goal is to train an object that, after applying to an input sequence $\vx$, can produce sequences $\vx^\prime$ that follow the probability distribution of the underlying \CA rules. 
As shown in \cite{HanZhang2018, GuoPoletti2018}, the probability amplitude of obtaining an arbitrary sequence $\vx$ can be computed from a special case of tensor networks, known as matrix product state (MPS)
\begin{align}
	P(\vx) &= \sum_{a_0, \dots, a_L} M^{x_{1}}_{a_0\;a_1}M^{x_{2}}_{a_1\;a_2}\dots M^{x_{L}}_{a_{L-1}\;a_L} =M^{\vx}.
    \label{eq:MPS_prob}
\end{align} 
Here, $M^{x_{l}}_{a_{l-1}\;a_l}$ is a three-legged tensor with one physical index $x_{l}$ and two auxiliary indices $a_{l-1}$ and $a_{l}$ which link the $l-$th tensor to the previous and the next one.
The number of possible elements in $a_l$ is called {\it bond dimension} which we refer to as $D$. For the MPS in Eq.~(\ref{eq:MPS_prob}) we use the short form $M^{\vx}$, and we note that it can be normalized such that $\sum_{\vx}P(\vx)=1$ \footnote{The computation of the norm, which needs to be used to rescale the MPS, can be implemented by preparing the $D=1$ state $\pmb{1} = \otimes_{l=1}^L I_l $ where $I_l=(1,1)_l$ is the identity vector at site $l$ and then multiplying the MPS $P(\vx)$ with $\pmb{1}$.}.    

In contrast, matrix product operators (MPOs) are linear operators $W^{\vx'}_{\vx}$ composed of four-legged tensors $W^{x_l\;x'_l}_{b_{l-1}\;b_l}$. We thus write 
\begin{align}\label{eq:MPO}
    W^{\vx'}_{\vx} = \sum_{b_0\; b_1\;\dots \;b_L} W^{x_1\;x'_1}_{b_0\;b_1}W^{x_2\;x'_2}_{b_1\;b_2}\dots W^{x_L\;x'_L}_{b_{L-1}\;b_L}.
\end{align}
In Eq.\ref{eq:MPO}, the size of the auxiliary index $b_l$ is referred to as the operator bond dimension and we denote it with $D_W$. 

\begin{figure}[ht]
	\includegraphics[width=\columnwidth]{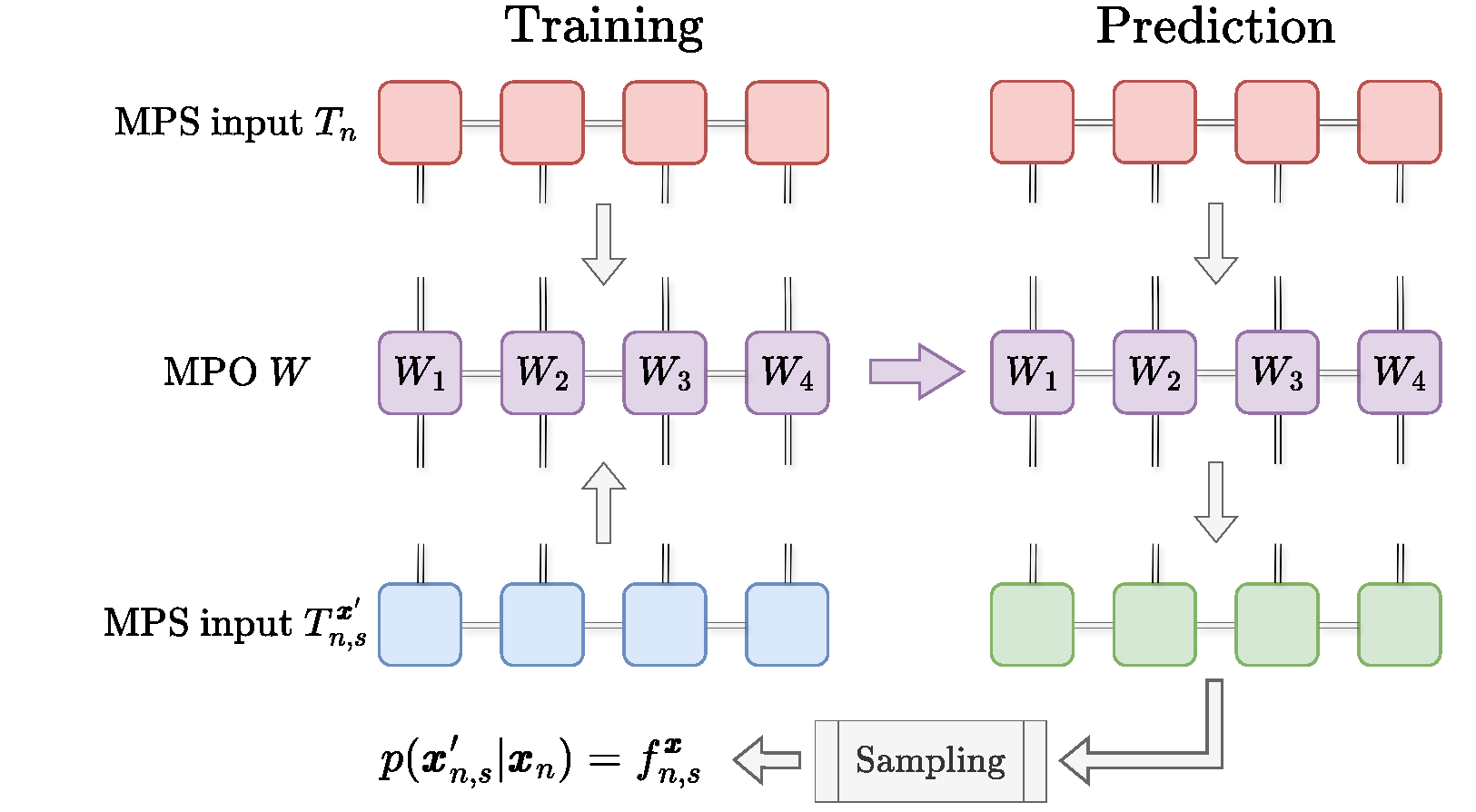}
\caption{Illustration of the probabilistic learning protocol. In the left portion, an input state and more than one output state are contracted with the MPO object which is in turn fed into the minimization protocol. In the right portion, we demonstrate the prediction, contracting an input state with the, now, trained, MPO, which is afterward subject to a sampling algorithm to recover a possible output with correct probability.}
	\label{fig1:graphical}
\end{figure}

Going back to the specific issue of sequences coming from \CA evolution, any input sequence $\vx$ or output sequence $\vx^\prime$ can be written as an MPS of bond dimension $D=1$ which we refer to as $T^{\vx}$ and $T^{\vx^\prime}$ \cite{SchollwockMPS}. When an MPO acts on an MPS, the result is still an MPS. 
We can thus write the conditional probability of obtaining $\vx^\prime$ from $\vx$ as  
\begin{align} 
    p(\vx^\prime|\vx) = T^{\vx^\prime\dagger} W^{\vx'}_{\vx} T^{\vx}.   
\end{align}  
Once learned, the MPO $W_{\vx}^{\vx'}$ can correctly predict the output sequences of the probabilistic \CA used in the training. A graphical representation of the learning process can be found in Fig.~\ref{fig1:graphical}. In this work, we consider the scenario in which $W_{\vx}^{\vx'}$ is time-translational invariant, meaning that the probability distribution of the \CA rules is the same at each time step.    

\section{Training}
\label{sec:training}
For the supervised learning, we need unique pairs of $\vx$ and $\vx^\prime$ and their corresponding conditional probability $p(\vx^\prime|\vx)$. We prepare the training data by randomly generating $N$ different $\vx$ and evolving them repeatedly with $S$ probabilistic \CA rules. We use the subscript $n$ to differentiate the input sequences $\vx_n$ and the subscript $s$ to differentiate the output sequences $\vx^\prime_{n,s}$ from $\vx_n$. At each time step, one input sequence $\vx_n$ can yield up to $S$ different output sequences $\vx^\prime_{n,s}$ (some input and output pairs may be the same for different rules). From this training data, we identify unique pairs ($\vx_n$, $\vx^\prime_{n,s}$) and their respective normalized conditional probability or frequency $p(\vx^\prime_{n,s}|\vx_n)=f^{\vx}_{n,s}$. The latter can be estimated by counting the occurrences of the different output sequences from the same input sequence
\begin{equation}\label{eq:freq}
    f^{\vx}_{n,s} = \frac{{\rm{count}} \left( x_n \to x^\prime_{n,s} \right)}{\sum_{s} {\rm{count}} \left(x_n \to x^\prime_{n,s}\right)}.
\end{equation}
In general, Eq.~\ref{eq:freq} can be used to approximate the conditional probability in any data set containing labeled input and outputs.

To learn the MPO $W_{\vx}^{\vx^\prime}$ from the training data, we first express the unique input $\vx_n$ and output sequences $\vx_{n,s}^\prime$ as MPS $T_{n}^{\vx}$ and $T_{n,s}^{\vx^\prime}$ respectively. 
With the frequency of the unique pairs $f^{\vx}_{n,s}$, we can construct a loss function from which to derive $W_{\vx}^{\vx^\prime}$. 
The main idea is that for any input sequence $\vx_n$ and all of its output sequences $\vx'_{n,s}$ we should get 
\begin{align}
    W_{\vx}^{\vx'} T^{\vx}_n = \sum_s \tilde{T}^{\vx'}_{n,s},   
\end{align} 
where $\tilde{T}^{\vx'}_{n,s} = f^{\vx}_{n,s} T^{\vx'}_{n,s}$, and this equation can be obtained by minimizing 
\begin{align}
    C_0 = \sum_n \left(W_{\vx}^{\vx'} T^{\vx}_n - \sum_s \tilde{T}^{\vx'}_{n,s}\right)^\dagger\left(W_{\vx}^{\vx'} T^{\vx}_n - \sum_s \tilde{T}^{\vx'}_{n,s}\right). \label{eq:c0}     
\end{align}
We further add a regularization term to $C_0$ that prevents overfitting so the overall loss function becomes 
\begin{align}
    C = C_0 + \alpha~ \Tr \left(W^{\vx^\prime \dagger}_{\vx} W^{\vx^\prime}_{\vx}\right), \label{eq:c}     
\end{align}
where the positive constant $\alpha$ is a regularization constant. For this work, we use $\alpha = 0.001$ as it was shown to provide good results \cite{GuoPoletti2018}. 

In general, one can minimize the loss function defined in Eq.~(\ref{eq:c}) using auto-differentiation and out-of-the-box optimizers like stochastic gradient descent or Adam. 
However, given the one-dimensional structure of the MPO, we can adopt a more efficient and effective approach. This approach is inspired by the ground state search algorithm in matrix product states~\cite{SchollwockMPS}. 

We locally and sequentially optimize each tensor $W^{x_l\;x'_{l}}_{b'_{l-1}\;b'_l}$ by solving a linear problem. 
In each optimization {\it sweep}, we update the tensors sequentially from site $1$ to site $L$, and then back to site $1$. This sweep is then repeated until the loss function converges or falls below a stopping criterion. For more details see App.~\ref{app:optimization}.  

\section{Prediction} 
\label{sec:prediction}   

Given a well trained MPO $W_{\vx}^{\vx'}$ and an input sequence $\vx_{n}$, we generate an MPS $M^{\vx^\prime | \vx_n}$ from which different output sequences $\vx^{\prime}_{n,s}$ can be extracted with the correct frequencies $f_{n,s}^{\vx^\prime}$. We use the zipper algorithm~\cite{HanZhang2018, PeresCasagrandePoletti2023} to extract these output sequences. Starting from the MPS
\begin{align}
    M^{\vx^\prime | \vx_n} = W_{\vx}^{\vx'} T_n^{\vx},
\end{align}
we first compute the normalized probability of finding the first site of the sequence $x_1 = 1$ from $M^{\vx^\prime | \vx_n}$,
\begin{align}
    m_1 = \frac{\Tr(\su_1 M^{\vx^\prime | \vx_n})}{\Tr(M^{\vx^\prime | \vx_n})},
\end{align}
where
\begin{align}
    \su_l = \left[\begin{array}{cc}
        1 & 0 \\
        0 & 0
    \end{array}\right], \; 
    \sigmad_l = \left[\begin{array}{cc}
        0 & 0 \\
        0 & 1
    \end{array}\right],
\end{align}
are the local spin-up and spin-down operators.
A random number $r$ is then drawn from a uniform distribution $U_{0,1}$ between $0$ to $1$. If $r < m_1$ then we set $x_1 = 1$. Otherwise, we set $x_1 = 0$. At site $l$ we thus have 
\begin{align}
    m_l = \frac{\Tr\left(\sigma^{a_1}_1\dots\sigma^{a_{l-1}}_{l-1}\su_l M^{\vx^\prime | \vx_n} \right) } { \Tr\left(\sigma^{a_1}_1\dots\sigma^{a_{l-1}}_{l-1} M^{\vx^\prime | \vx_n} \right)  },
\end{align}
where $a_{l}= u$ or $d$ depending on the $1$ or $0$ we set on the previous sites.
Similarly, we compare $m_l$ with a random number $r$ drawn from $U_{0,1}$ and set $x_l$ to be either $1$ ($r \leq m_1$) or $0$ ($r > m_1$).

\section{Characterization}\label{sec:characterization} 

We consider the case of a finite number $S$ of different rules, and each rule $s$ with an expected probability $p^e_s$. 
To evaluate the predicted probability of rule $s$, $p^p_s$, we prepare $N$ testing input sequences $\vx_n$ and extract one output sequence $\vx^\prime_{n,s}$ from each $\vx_n$. From these $N$ pairs of input-output sequences, we evaluate the frequency of a rule $s$.
The error of our predictions is given by 
\begin{equation}
    \epsilon = \dfrac{1}{S} \sum_{s = 1}^{S} |p^{p}_s - p^{e}_s|. 
    \label{eq:error}
\end{equation}
Occasionally, different rules can produce the same output sequence from an input sequence. This scenario is removed only during the characterization of performance so that Eq.~(\ref{eq:error}) remains a valid indicator.

\section{Analytical considerations}\label{sec:analytical} 
It was shown that the dynamics stemming from deterministic \CA can be described exactly with MPO~\cite{GuoPoletti2018} and each of the 256 local rules can be represented by an MPO of bond dimensions $D_W \leq 4$. This analytical result can be extended further to probabilistic \CA . To understand how to write exact MPO solutions for probabilistic \CA , we summarize here the key points on how to generate the MPO solution for non-probabilistic \CA . 
In the typical \CA where the same rule is applied at each step, there is a one-to-one mapping between the input and output sequences. Hence, for the MPO $W_{\vx}^{\vx'}$, if one uses the sequences of $\vx_n$ and $\vx^\prime_{n,s}$ to evaluate $W_{\vx}^{\vx^\prime}$, the MPO should be $1$ for $\vx^\prime_{n,s}$ and $0$ for all other possible output sequences $\vx'$. For example, for a system with rule $30$, the input sequence $0,0,1,0,0$ becomes the output sequence $0,1,1,1,0$ and thus $W^{01110}_{00100}=1$. All other combinations of the upper indices give $0$. 
Since $W_{\vx}^{\vx'}$ is the product of tensors, we write the tensors $W^{x_{l}\; x'_{l}}_{b_{l-1}\;b_{l}}$ such that their products give $1$ only when the input and the output sequences are connected by the relevant \CA rule. 
The required bond dimension is $D_W=1$ if the rule depends only on the local site (e.g. rule $51$), $D_W=2$ if the rule depends on the local site and one of the nearest neighbors (e.g. rule $153$), and $D_W=4$ if the rule depends on the local site and both nearest neighbors (e.g. rules $18$ or $30$).     

We show that probabilistic \CA can be written exactly with MPOs as well. 
To build this MPO, we multiply the MPO for each rule by its corresponding probability and then add them together. The bond dimension of the resultant MPO is thus the sum of the $S$ constituents. 
We note that this analytical solution is not unique because (i) one can perform gauge transformations on the tensors by including matrices between the tensors, and (ii) this analytical MPO may be further compressed to be described exactly by a smaller bond dimension.

\section{Results}\label{sec:results}  
In this section, we show the prediction performance for various probabilistic \CA.
For most of the results, the number of probabilistic rules is $S=2$ and the training is done with $N=20000$ training inputs of length $L=20$ unless otherwise stated. After training, we consider $N=10000$ different input sequences $\vx_n$ for the characterization. We use the zipper algorithm in Sec.~\ref{sec:prediction} to sample one output sequence $\vx^\prime_{n,s}$ from each of these inputs. We then check if the input-output pair follows the first rule, the second rule, or none of them. By counting how many times the output sequences have followed one of the two rules, we can check if these frequencies correspond to the expected conditional probabilities. Naturally, the frequency should be $0$ for all the other rules. 
We note that for a system of size $20$, the full sample space would be $2^{20}$, or around one million states. Our training set of $N=20000$ is therefore just $\sim 2\%$ of the total configurations. 

\subsection{Performance of MPO learning model}

\begin{figure}[ht]
    \includegraphics[width=\columnwidth]{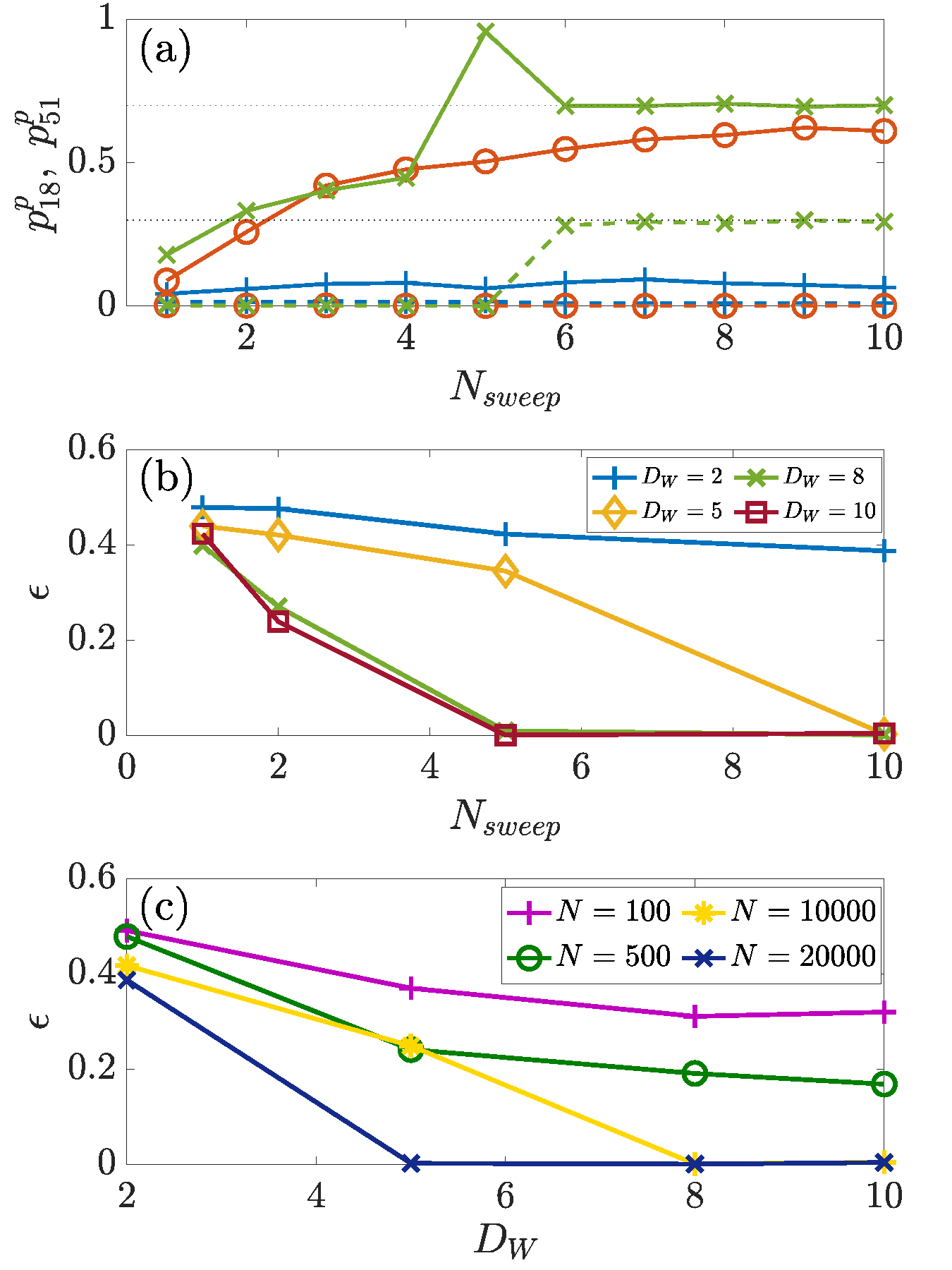}
 	\caption{MPO training for probabilistic \CA with rules 18 and 51, with $p_{18}^e = 0.7$ and $p_{51}^e = 0.3$ respectively. (a) Predicted probabilities $P^p_{18}$ and $p_{51}^p$ against the number of training sweeps. Different symbols present different MPO bond dimensions: $D_W = 2$ is shown in blue ``+'', $D_W = 4$ is shown in orange circles, and $D_W = 8$ is shown in green ``x''. The solid lines are for $p^p_{18}$ and the dashed lines are for $p^p_{51}$. The dotted horizontal line marks the exact probabilities $p^e_{18}$ and $p_{51}^e$. (b) Prediction error $\epsilon$ against the number of training sweeps. The different lines represent different $D_W$. (c) $\epsilon$ against different $D_W$. The different lines represent different training samples $N$.
    The parameters used are system size $L = 20$, number of samples $N=20000$, and number of sweeps $N_{\text{sweep}}=10$, unless otherwise specified in the subplot descriptions.}
  	\label{fig3-18vs51}
\end{figure}

We first investigate the performance of the MPO learning model by analyzing the effects of changing the number of training sweeps $N_{sweep}$, MPO bond dimension $D_W$, and the number of training samples $N$. 
In Fig.~\ref{fig3-18vs51} we consider the case in which the two probabilistic rules are $18$ (periodic with a large period) and $51$ (periodic with period 2), with probabilities $p_{18}^e$ and $p_{51}^e=1-p_{18}^e$ respectively.
In Fig.~\ref{fig3-18vs51}(a), we plot the predicted probabilities $p_{18}^p$ (solid lines) and $p_{51}^p$ (dashed lines) against the number of training sweeps $N_{sweep}$. The exact probabilities are shown as the dotted lines, with $p_{18}^e = 0.7$ and $p_{51}^e = 0.3$. 
We find that $D_W$ plays an important role in identifying the right rules and their corresponding probabilities. For this combination of rules, a bond dimension $D_W=5$ is sufficient to recover the exact probabilities. Further increasing $D_W$ extends the generalizing power of the MPO, and therefore improves the accuracy of the results. This is also illustrated in Fig.~\ref{fig3-18vs51}(b), in which we plot the error $\epsilon$ against the number of sweeps. When $D_W$ is larger, $p^p_{18}$ converges to $p^e_{18}$ after a smaller number of sweeps. On the other hand, $D_W = 2$ does not produce the right $p^p_{18}$ despite increasing the number of sweeps. Lastly, in Fig.~\ref{fig3-18vs51}(c), we plot the error $\epsilon$ against $D_W$ for different number of training samples $N$. Although $D_W = 5$ is theoretically sufficient to give the right predictions, a minimum $N=20000$ is required to achieve accurate convergence at $N_{sweep} = 10$. For smaller $N$, the error can be mitigated by increasing $D_W$.

\begin{figure}[ht]
    \includegraphics[width=\columnwidth]{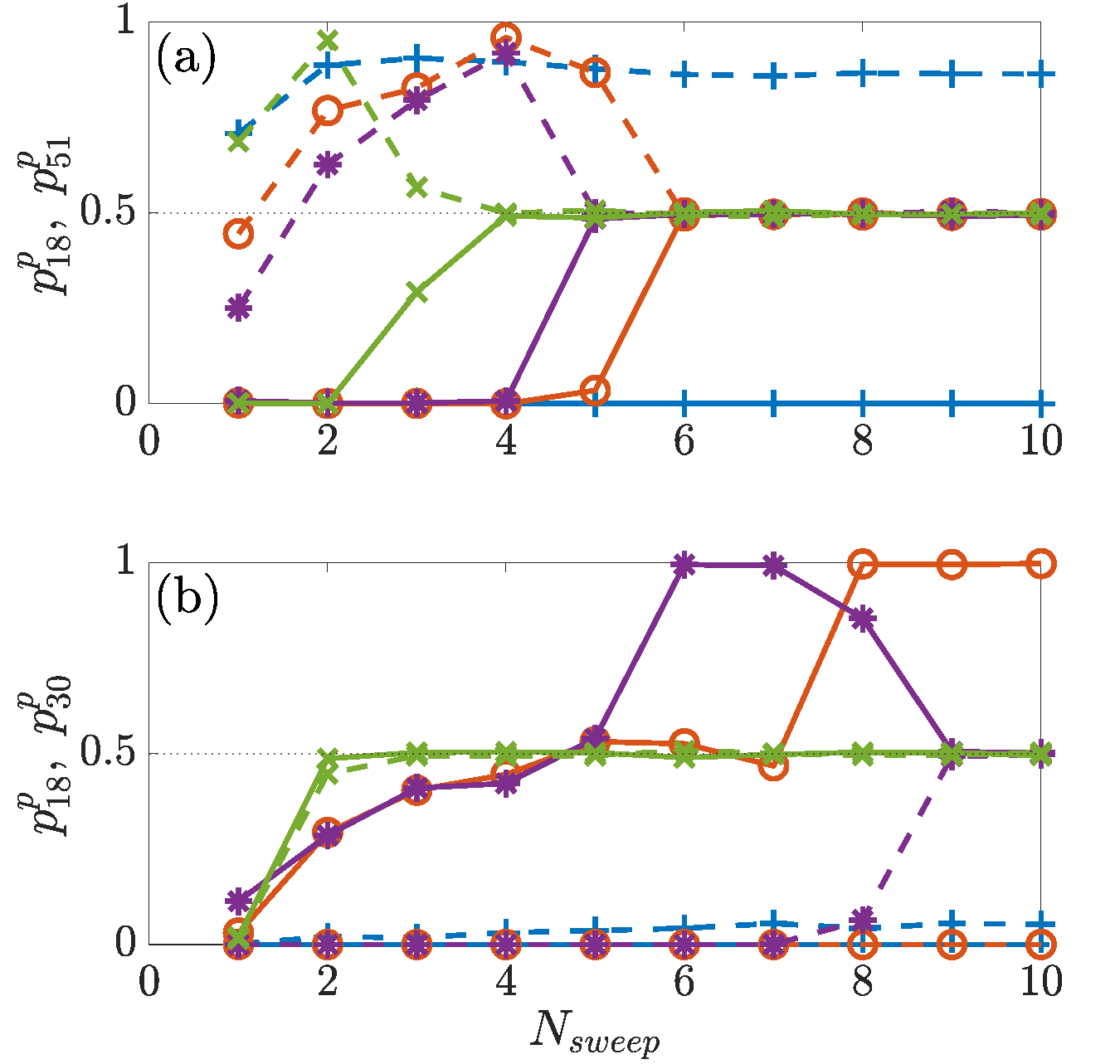}
	\caption{MPO training for probabilistic \CA with different pairs of rules. For each rule, the probability of occurrence is $p^e_i = 0.5$. The different markers represent different MPO bond dimensions: blue ``+'' for $D_W = 2$, orange circles for $D_W = 4$, purple asterisks for $D_W = 6$, and green ``x'' for $D_W = 8$. The solid and dashed lines correspond to the predicted probabilities of different rules. The dotted lines correspond to the exact probabilities. (a) Predicted probabilities $p^p_{18}$ and $p_{51}^p$ against the number of training sweeps. (b) Predicted probabilities $p^p_{18}$ and $p_{30}^p$ against the number of training sweeps. 
    The parameters used are system size $L = 20$ and number of samples $N=20000$.}
	\label{fig4-differentRules_t-0.5}
\end{figure}

In Fig.~\ref{fig4-differentRules_t-0.5}, we show that our insights are generic by comparing the results of two different pairs of probabilistic rules. In Fig.~\ref{fig4-differentRules_t-0.5}(a) we consider the same scenario with rules $18$ and $51$, but with $p^e_{18} = p^e_{51} = 0.5$. For this scenario, $D_W=5$ is sufficient to guarantee accurately predicted probabilities.  In Fig.~\ref{fig4-differentRules_t-0.5}(b) we consider a different scenario with rules $18$ and $30$. For this pair of probabilistic rules, we expect and observe that $D_W=8$ is sufficient to provide fast and accurate predictions.
In both cases, increasing $D_W$ beyond the analytical consideration leads to the prediction of the correct probabilities.

\begin{figure}[ht]
    \includegraphics[width=\columnwidth]{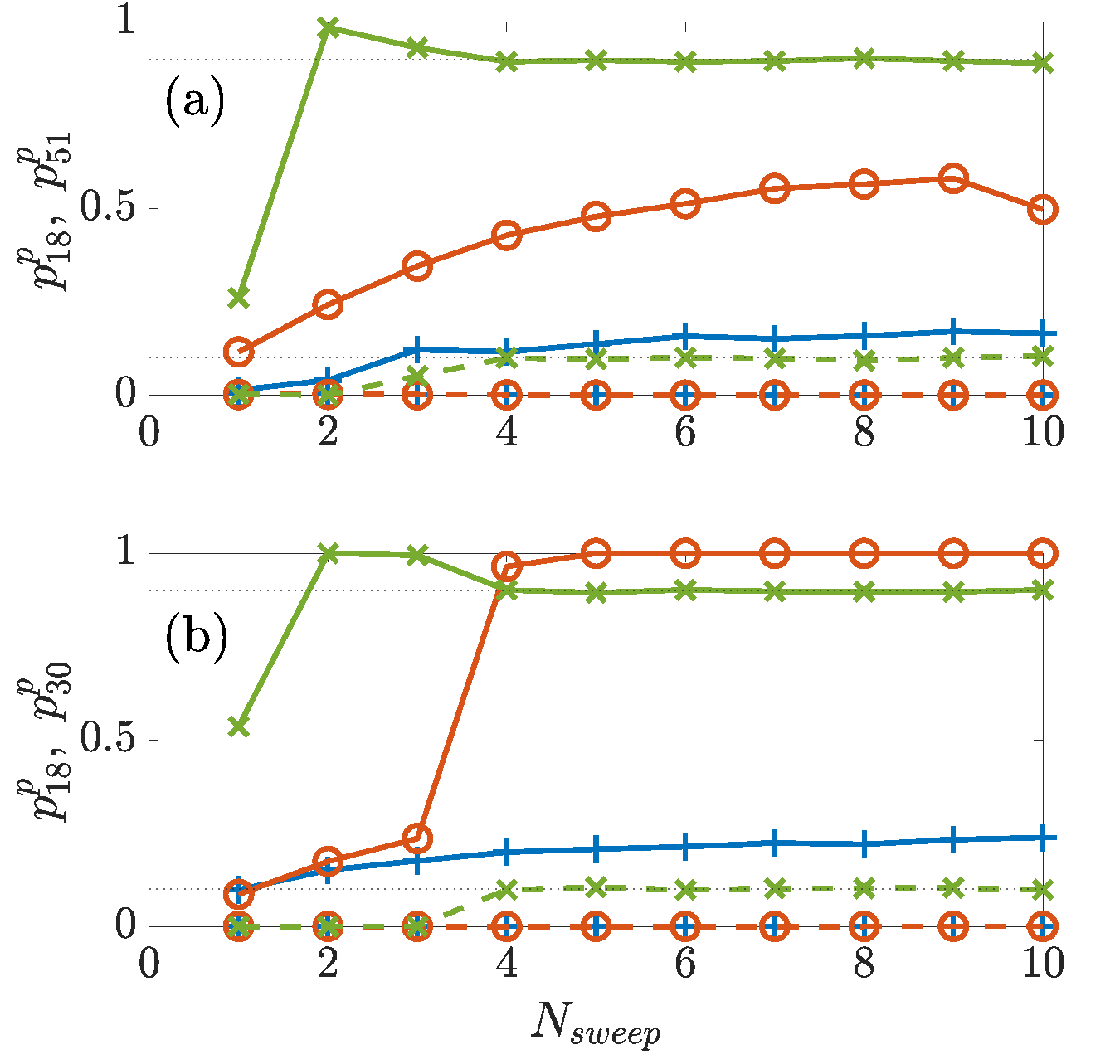}
	\caption{MPO training for probabilistic \CA with different pairs of rules. The probabilities of the rules are $p^e_{18} = 0.9$ and $p^e_{51/30} = 0.1$. Different markers correspond to different bond dimensions: blue ``+'' for $D_W = 2$, orange circles for $D_W = 4$, and green ``x'' for $D_W = 8$. The solid and dashed lines correspond to the predicted probabilities of different rules. The dotted lines correspond to the exact probabilities. (a) Predicted probabilities $p^p_{18}$ and $p_{51}^p$ against the number of training sweeps. (b) Predicted probabilities $p^p_{18}$ and $p_{30}^p$ against the number of training sweeps.  The parameters used are system size $L = 20$ and number of samples $N=20000$.}
	\label{fig5-differentRules_t-0.9}
\end{figure}

When $p^e_i$ deviates from $0.5$, the MPO training becomes more difficult. If the probability of one rule is considerably larger than the other, the optimization algorithm could be trapped in a local minimum, resulting in the prediction of a deterministic \CA. In this case, the second rule is not regarded as a characteristic of the probabilistic \CA, but as noise. Despite the increased difficulty, our algorithm can distinguish the two probabilistic rules satisfactorily. To demonstrate this, we present the same two pairings of rules shown in Fig.~\ref{fig4-differentRules_t-0.5}, but with $p_{18}^p = 0.9$ and $p_{51/30} = 0.1$. The predicted probabilities are plotted against the number of training sweeps in Fig.~\ref{fig5-differentRules_t-0.9}. In both scenarios, we find that an MPO with $D_W$ greater than or equal to the analytical consideration can predict the correct probabilistic \CA dynamics.

\begin{figure}[ht]
     \includegraphics[width=\columnwidth]{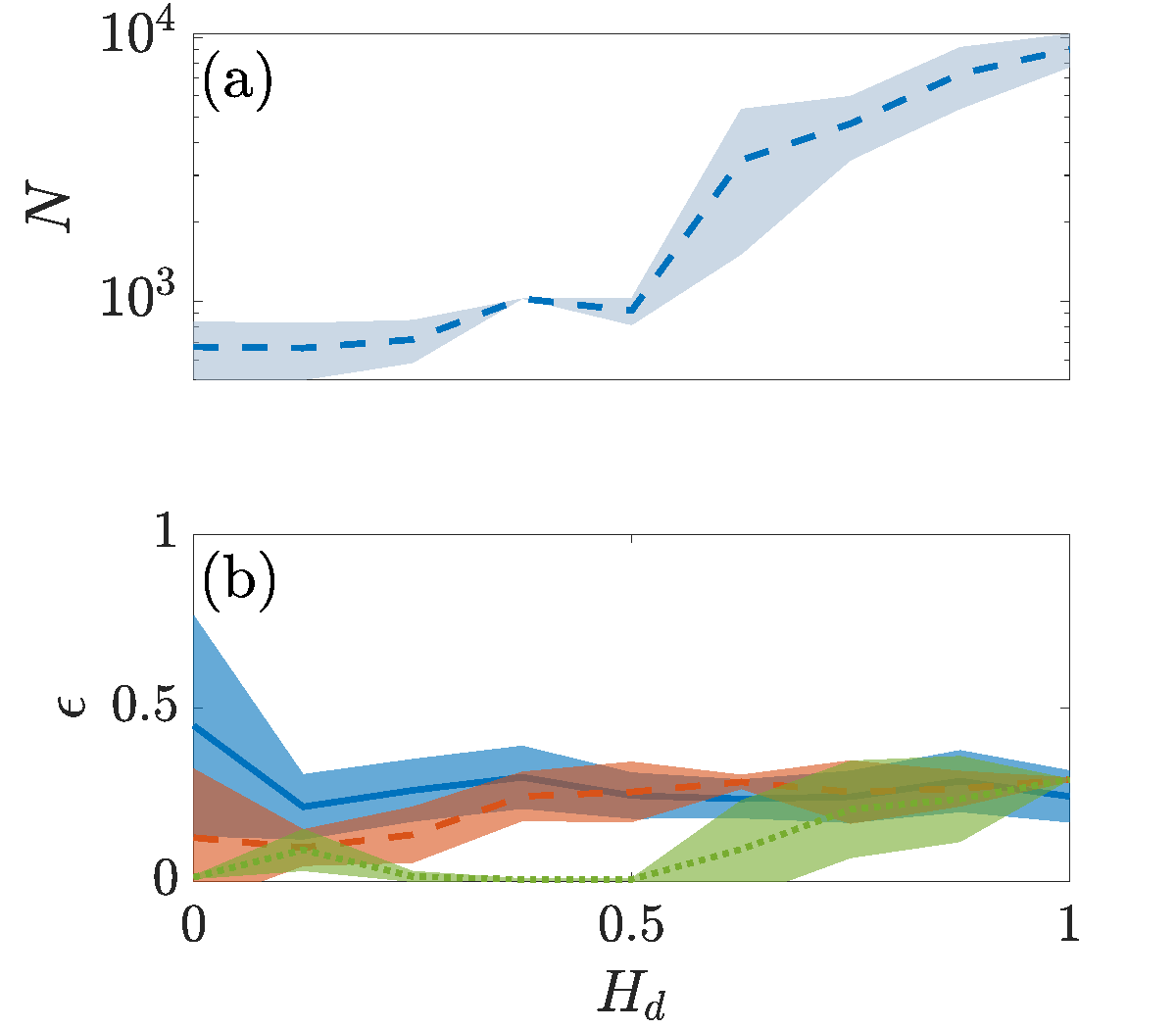}
	\caption{(a) Average number of training samples required to obtain prediction error $\epsilon < 0.05$ plotted against the bit-wise Hamming distance $H_d$ between the rules. The number of training sweeps is $20$. (b) Prediction error $\epsilon$ plotted against the bit-wise Hamming distance $H_d$ between two rules. The blue continuous line corresponds to one training sweep, the orange dashed line is for 5 training sweeps, and the green dotted line is for 20 sweeps. The number of training samples is $N = 2048$. In both panels, for each Hamming distance, we average the $\epsilon$ of ten pairs of rules chosen randomly, with one rule having a probability of $0.7$ and the other $0.3$. The lines represent the average $\epsilon$ of the ten pairs and the shaded region is the standard deviation. Other parameters used are bond dimension $D_W = 8$ and system size $L=14$. }
	\label{fig6-errorVsSampleSizeAndHammingDistance}
\end{figure}

Another factor that affects the difficulty of MPO training is the distance between the probabilistic rules. In the bit-wise representation of \CA~\cite{Wolfram1983}, the distance between rules can be characterized by their Hamming distance $H_d$. 
For clarity, we provide an example of the bit-wise distance between two rules: rule $18$ is $00010010$ bit-wise, and rule $30$ is $00011110$. Their $H_d$ is $1/4$ because two bits out of eight are different.   
Given a particular $H_d$, we generate ten different random pairs of rules and analyze the performance of our MPO training algorithm.
For the ten different pairs of rules studied in Fig.~\ref{fig6-errorVsSampleSizeAndHammingDistance}, the rules have probabilities $0.7$ and $0.3$ and the system size is $L=14$.
In Fig.~\ref{fig6-errorVsSampleSizeAndHammingDistance}(a) we plot the number of training samples $N$ needed to reduce the error of the prediction to $\epsilon \leq 0.05$. We observe a significant increase in $N$ for $H_d > 0.5$, suggesting that rules that are more different from each other are harder to learn. 
To corroborate this, we plot the prediction error $\epsilon$ against $H_d$ for different numbers of training sweeps in Fig.~\ref{fig6-errorVsSampleSizeAndHammingDistance}(b). In this panel, the lines give the average $\epsilon$ of the ten different pairs of rules and the shaded region represents the standard deviation. The MPO bond dimension is $D_W = 8$ and the number of training samples is $N=2048$. For a single training sweep, e.g. the blue continuous line, $\epsilon$ is consistently large for all $H_d$. As the number of training sweeps increases, e.g. the green dotted line (20 sweeps), rules with a $H_d < 0.5$ can be predicted with smaller errors than those with larger $H_d$. 
 
\begin{figure}[ht]
    \includegraphics[width=\columnwidth]{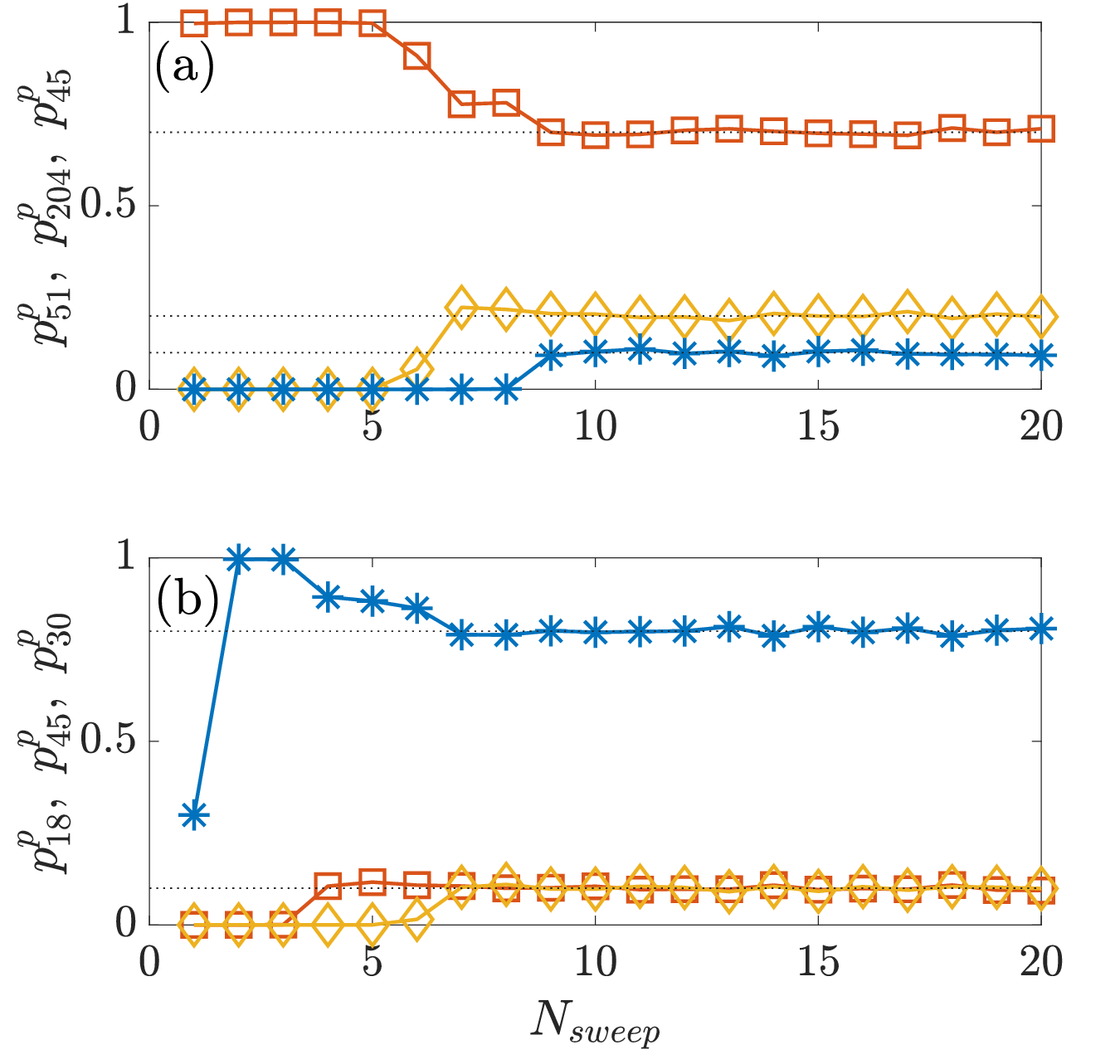}
	\caption{MPO training for probabilistic \CA with three rules. (a) The predicted probabilities of rules $51$ (blue asterisks), $204$ (orange squares), and $45$ (yellow diamonds) against the number of training sweeps. The dotted lines represent the exact probabilities $p_{51}^e = 0.1$, $p_{204}^e=0.7$, and $p_{45}^E=0.2$. The MPO bond dimension is $D_W=8$. (b) The predicted probabilities of rules $18$ (blue asterisks), $30$ (orange squares), and $45$ (yellow diamonds) against the number of training sweeps. The dotted lines represent the exact probabilities $p_{18}^e=0.8$, and $p_{30}^e=p_{45}^e=0.1$. The MPO bond dimension is $D_W=10$. In all panels, the number of training samples is $N=20000$ and the system size is $L=20$. }
	\label{fig7-ThreeRulesCA}
\end{figure}

Lastly, to exemplify the generality of our approach, we consider the case in which three probabilistic rules are present.
In Fig.~\ref{fig7-ThreeRulesCA} we show the predicted probabilities of the rules against the number of training sweeps. In panel (a), the rules are $51$, $204$, and $45$ with probabilities $p^e_{51} = 0.1, p^e_{204} = 0.7, p^e_{45} = 0.2$ respectively. In panel (b), the rules are $18$, $30$, and $45$ with probabilities $p^e_{18} = 0.8, p^e_{30} = p^e_{45} = 0.1$ respectively.
We show clearly that our MPO training algorithm can accurately learn the dynamics with three probabilistic rules, even if the probability of one rule is much higher than the others. 

\section{Conclusions}\label{sec:conclusions} 
We have extended the matrix product operator sequence-to-sequence learning algorithm to study probabilistic \CA dynamics. Our new MPO learning algorithm uses a different protocol to collect and analyze the training data. Most importantly, we introduce a loss function that considers more than one possible output for a given input sequence and implement a different approach to sample the output sequence from the trained model. Given the probabilistic nature of the problem we study, we use a different indicator to characterize the performance of the trained model \cite{GuoPoletti2018}.
We show that, with small MPO bond dimensions and training samples, it is possible to learn probabilistic \CA sequences both when the rules appear with very different probabilities and when their probabilities are similar. The chaotic or regular nature of the rules does not play a significant role in the performance of the algorithm. For the cases we have studied, the predicted probabilities always converge toward the correct distribution after a small number of training sweeps. 

In future works, we could test the performance of this learning model in extracting evolution rules from systems with more complex probability distributions. We also plan to investigate the performance with non-spatially translational rules and consider rules for which the evolution is identical for a large number of input sequences. 

\begin{acknowledgments}
H.P.C acknowledges Pedro Toledo and Ivan Iakoupov's valuable input and discussions on this work. 
D.P. and B. X. acknowledge support from the Ministry of Education Singapore, under the grant MOE-T2EP50120-0019, from joint Israel-Singapore NRF-ISF Research grant NRF2020-NRF-ISF004-3528, and from the National Research Foundation, Singapore and the Agency for Science, Technology and Research (A*STAR) under the Quantum Engineering Programme (NRF2021-QEP2-02-P03). 
The computational work for this article was partially performed at the National Supercomputing Centre, Singapore \cite{NSCC}.
The codes and data generated are both available upon reasonable request to the authors.

\end{acknowledgments}

\begin{appendix}
\section{Optimization algorithm} \label{app:optimization}

For the optimization, we rely on the linear character of the loss functions in Eq.~(\ref{eq:c}). Taking the derivative over $W^{x_l\;x'_{l}}_{b'_{l-1}\;b'_l}$ and setting it to zero allows us to locally improve the tensor $W^{x_l\;x'_{l}}_{b_{l-1}\;b_l}$ of the MPO by solving the linear equation     
\begin{align}
    \mathcal{A}_{x_l\;b_{l-1}\;b_l}^{y_l\;b'_{l-1}\;b'_l} \; W^{x_l\;x'_l}_{b_{l-1}\;b_l} = \mathcal{B}^{y_l\;x'_l}_{b'_{l-1}\;b'_l}. 
\end{align} 
Here $\mathcal{A}_{x_l\;b_{l-1}\;b_l}^{y_l\;b'_{l-1}\;b'_l}$ and $\mathcal{B}^{y_l\;x'_l}_{b'_{l-1}\;b'_l}$ are given by 
\begin{align}
\mathcal{A}_{x_l\;b_{l-1}\;b_l}^{y_l\;b'_{l-1}\;b'_l} = & \sum_{n} \left( L^{n}_{b_{l-1}\;b'_{l-1}} R^{n}_{b_l\;b'_l}T_n^{x_l}T_{n}^{y_l}   \right) \nonumber \\ 
&  +  \alpha~ \mathcal{C}_{b_{l-1}\;b'_{l-1}} \mathcal{D}_{b_{l}\;b'_{l}}  
\end{align} 
with 
\begin{align}
 L^{n}_{b_{l}\;b'_{l}} = L^{n}_{b_{l-1}\;b'_{l-1}} W^{x_l\;x'_{l}}_{b_{l-1}\;b_l} W^{y_l\;y'_{l}}_{b'_{l-1}\;b'_l} T_n^{x_l} T_{n}^{y_l} \delta_{x'_l,y'_l},    
\end{align} 
\begin{align}
 R^{n}_{b_{l-1}\;b'_{l-1}} = R^{n}_{b_{l}\;b'_{l}} W^{x_l\;x'_{l}}_{b_{l-1}\;b_l} W^{y_l\;y'_{l}}_{b'_{l-1}\;b'_l} T_n^{x_l} T_{n}^{y_l}\delta_{x'_l,y'_l},    
\end{align} 
\begin{align}
\mathcal{C}_{b_{l-1}\;b'_{l-1}} = \mathcal{C}_{b_{l-2}\;b'_{l-2}} W^{x_{l-1}\;x'_{l-1}}_{b_{l-2}\;b_{l-1}} W^{x_{l-1}\;x'_{l-1}}_{b'_{l-2}\;b'_{l-1}},  
\end{align} 
and 
\begin{align}
\mathcal{D}_{b_{l}\;b'_{l}} = \mathcal{D}_{b_{l+1}\;b'_{l+1}} W^{x_{l+1}\;x'_{l+1}}_{b_{l}\;b_{l+1}} W^{x_{l+1}\;x'_{l+1}}_{b'_{l}\;b'_{l+1}}    , 
\end{align} 
while 
\begin{align}
\mathcal{B}^{y_l\;x'_l}_{b'_{l-1}\;b'_l} = \sum_{n,s} A^{n,s}_{b'_{l-1}} \tilde{T}^{x'_l}_{n,m} T^{y_l}_{n}  B^{n,s}_{b'_l}
\end{align} 
with 
\begin{align}
A^{n,s}_{b'_{l}} = A^{n,s}_{b'_{l-1}} W^{y_{l}\;y'_{l}}_{b'_{l-1}\;b'_{l}} \tilde{T}^{x'_l}_{n,s} T^{y_l}_{n}\delta_{x'_l,y'_l},      
\end{align} 
and 
\begin{align}
B^{n,s}_{b'_{l-1}} = B^{n,s}_{b'_{l}} W^{y_{l}\;y'_{l}}_{b'_{l-1}\;b'_{l}} \tilde{T}^{x'_l}_{n,s} T^{y_l}_{n}\delta_{x'_l,y'_l},  
\end{align} 
In the above equations, summation over common indices is assumed, and $\delta_{a,b}$ is the Kronecker delta. 

Once one solves for $W^{x_l\;y_l}_{b_{l-1}\;b_l}$, then new $\mathcal{A}_{x_{l+1}\;b_{l}\;b_{l+1}}^{x'_{l+1}\;b'_{l}\;b'_{l+1}}$ and $\mathcal{B}^{x'_{l+1}\;y'_{l+1}}_{b'_{l}\;b'_{l+1}}$ can be generated (or $\mathcal{A}_{x_{l-1}\;b_{l-2}\;b_{l-1}}^{x'_{l-1}\;b'_{l-2}\;b'_{l-1}}$ and $\mathcal{B}^{x'_{l-1}\;y'_{l-1}}_{b'_{l-2}\;b'_{l-1}}$ depending on what is the next tensor to be optimized). This is done sequentially, ``sweeping'' through all the site $l$ from $1$ to $L$ and back to $1$.

\end{appendix}
\bibliography{refs} 
\end{document}